\documentclass[12pt,a4paper]{article}
\pdfoutput=1

\usepackage{epsfig}
\usepackage{amsmath}
\usepackage{amssymb}
\usepackage{verbatim}
\usepackage{bm}
\usepackage{dsfont}
\usepackage[footnotesize]{caption}
\usepackage{subfigure}
\usepackage{cancel}
\usepackage{cite}
\newcommand{\Mp}{M_{\mathrm{P}}}

\begin{document}

\begin{flushleft}
DESY 12-149\\
October 2012
\end{flushleft}

\vskip 1cm

\begin{center}
{\LARGE\bf Superconformal D-Term Inflation}

\vskip 2cm

{\large W.~Buchm\"uller, V.~Domcke,  K.~Schmitz}\\[3mm]
{\it{
Deutsches Elektronen-Synchrotron DESY, 22607 Hamburg, Germany}
}
\end{center}

\vskip 1cm

\begin{abstract}
\noindent 
We study models of hybrid inflation in the framework of supergravity
with superconformal matter. F-term hybrid inflation is not viable since the inflaton
acquires a large tachyonic mass. 
On the contrary, D-term hybrid inflation can successfully account for
the amplitude of the primordial power spectrum.
It is a two-field inflation model which, depending on parameters,
yields values of the scalar spectral index down to
$n_s \simeq 0.96$. Generically, there is a tension between a small spectral
index and the cosmic string bound albeit,
within $2\sigma$ uncertainty, the current observational bounds 
can be simultaneously fulfilled.
\end{abstract}.

\thispagestyle{empty}

\newpage

\section{Introduction}

Hybrid inflation \cite{Linde:1991km} is an attractive mechanism for generating 
the cosmological density perturbations. It is naturally realized in the
framework of grand unified theories (GUTs) and string theories, especially 
in the form of D-term inflation \cite{Binetruy:1996xj,Halyo:1996pp} where the 
GUT scale emerges via the Fayet-Iliopoulos (FI) term of an anomalous 
$U(1)$ symmetry. A further important virtue of D-term inflation is
that tree-level
supergravity corrections to the inflaton mass of order the Hubble parameter 
are absent.

D-term inflation has been quantitatively analyzed for the canonical
K\"ahler potential as well as for some non-minimal K\"ahler potentials 
\cite{Rocher:2004et,Battye:2006pk,Rocher:2006nh}. The value of the inflaton 
field is typically $\mathcal{O}(M_{\rm P})$ and supergravity corrections
are therefore important. In addition to the primordial fluctuations of
the inflaton field, the cosmic microwave background is significantly affected
by the production of cosmic strings at the end of inflation
\cite{Hindmarsh:1994re}. Generically, it appears difficult to obtain agreement 
with observational data \cite{Battye:2010hg}. In particular the scalar 
spectral index
$n_s$ turns out to be rather large and the gauge coupling is constrained to 
small values, in conflict with the motivation of implementing D-term 
hybrid inflation in GUTs.

In this paper we study D-term inflation in the context of
superconformal supergravity models \cite{Kallosh:2000ve} which have
recently been considered  in connection with Higgs inflation
\cite{Bezrukov:2007ep,Einhorn:2009bh,Ferrara:2010yw}. These models are
motivated by the underlying superconformal symmetry of supergravity,
and have 
several intriguing features.
In particular, there is a Jordan frame in which the matter part of the
Lagrangian takes a particularly simple form, closely resembling global
supersymmetry. In the Einstein frame, supergravity corrections to
scalar masses are suppressed by powers of $1/M_P$,
and, contrary to canonical supergravity, the scalar potential does not contain
factors which grow exponentially at large field values.
The superconformal symmetry is broken 
by fixing the value of the conformal compensator field,
which generates the kinetic term of the graviton
\cite{Kallosh:2000ve}. As we shall see, a Fayet-Iliopoulos term can be
introduced analogously.
A further explicit breaking of superconformal symmetry is a holomorphic 
contribution to the K\"ahler potential \cite{Ferrara:2010yw}. This turns D-term inflation into a two-field inflation model.   

As we have recently shown, the spontaneous breaking of $B$$-$$L$, the
difference of baryon and lepton number, at the GUT scale can explain the
initial conditions of the hot early universe including baryogenesis and
dark matter \cite{Buchmuller:2012wn,Buchmuller:2012bt}. This analysis was carried out assuming F-term hybrid inflation. As we shall see, F-term hybrid inflation is
inconsistent with superconformal symmetry. On the contrary, D-term inflation can be implemented with superconformal symmetry and
can also incorporate spontaneous $B$$-$$L$ breaking at the GUT scale.

The paper is organized as follows. In Section~\ref{sec_superconformal_inflation} we review the most important features of superconformal models of inflation and in particular discuss the resulting scalar potentials for F- and D-term hybrid inflation. Section~\ref{sec_single_field} deals with an important special case, namely a 
single-field scenario which arises if inflation lasted sufficiently
long before the onset of the final 50 e-folds. The full two-field inflation model is discussed in Section~\ref{sec_two-field}.
Our conclusions are presented in Section~\ref{sec_conclusion}.

\section{Superconformal models of inflation \label{sec_superconformal_inflation}}

\subsection{Supergravity with superconformal matter}

An attractive class of supergravity models can be defined by requiring
the matter sector and its couplings to supergravity to be invariant under
superconformal transformations. The superconformal symmetry is explicitly 
broken only by the pure supergravity part of the action and the superconformal 
anomaly. Matter interactions at energies below the Planck mass then obey the 
superconformal symmetry up to
corrections suppressed by inverse powers of the Planck mass and radiative
corrections \cite{Ferrara:2010yw}. 

For these theories there exists a
Jordan frame in which the Lagrangian takes a remarkably simple form which closely
resembles globally supersymmetric theories. The bosonic part for metric
and scalar fields $z^{\alpha}$ is given by \cite{Ferrara:2010yw}
\begin{align}\label{LagJ}
\frac{1}{\sqrt{-g_J}}\mathcal{L}_J = 
\frac{1}{2}\Mp^2 &\left(R_J + 6\mathcal{A}_{\mu}\mathcal{A}^{\mu}\right) - 
\frac{1}{6}|z|^2 R_J \nonumber\\
&- G_{\alpha\bar{\alpha}}g_J^{\mu\nu}
\widetilde{\nabla}_{\mu}z^{\alpha}\widetilde{\nabla}_{\nu}\bar{z}^{\bar{\alpha}}  -
g\partial_{\alpha}\Phi\left(T^a z\right)^{\alpha} D_a  - V_J \ .
\end{align} 
Here the subscript $J$ indicates quantities in the Jordan frame,
$|z|^2 = \delta_{\alpha \bar{\alpha}}z^{\alpha}\bar{z}^{\bar{\alpha}}$, and 
$G_{\alpha\bar{\alpha}} = \partial_{\alpha}\partial_{\bar{\alpha}}\Phi$, with 
$\partial_{\alpha} = \partial/\partial z^{\alpha}$ and
$\partial_{\bar\alpha} = \partial/\partial \bar{z}^{\bar\alpha}$ 
acting on the so-called 
frame function
\begin{align}\label{frame}
\Phi(z,\bar{z}) = -3\Mp^2 + |z|^2 \ , 
\end{align}
which is the coefficient function of the curvature scalar $R_J$ in 
Eq.~(\ref{LagJ}). The covariant derivative
$\widetilde{\nabla}_{\mu} = \partial_{\mu} - i\mathcal{A}_{\mu} -igA^a_{\mu}T^a$
contains an auxiliary gauge field $\mathcal{A}_{\mu}$ and the dynamical 
gauge fields $A^a_{\mu}$, with
$T^a$ being the corresponding generators; $D_a$ are the auxiliary
components of the vector superfields. The scalar potential is
determined by the superpotential $W$ and the gauge kinetic function $f$,
\begin{align}\label{PotJ}
V_J = G^{\alpha\bar{\alpha}} \partial_{\alpha}W \partial_{\bar{\alpha}}\overline{W}
+ \frac{1}{2}({\rm Re}f)_{ab}D_aD_b\ ,
\end{align}
with  $G^{\alpha\bar{\alpha}} = G^{-1}_{\alpha\bar{\alpha}}$; for the
frame function (\ref{frame}) one has 
$G^{\alpha\bar{\alpha}} = \delta^{\alpha\bar{\alpha}}$.
Local Weyl invariance requires a cubic superpotential $W$ and a constant 
gauge kinetic function $f$. It is remarkable that,
up to corrections described by the auxiliary field $\mathcal{A}_{\mu}$,
the matter part of the Lagrangian (\ref{LagJ}) is that of global
supersymmetry \cite{Ferrara:2010yw}.

The Lagrangian in the Einstein frame with metric $g$ is obtained by
performing the transformation
\begin{align} \label{eq_gJE}
g_{J\mu\nu} = \Omega^2 g_{\mu\nu}\  , \qquad \mathrm{with} \quad 
\Omega^2 = -\frac{3\Mp^2}{\Phi} 
= \left(1-\frac{|z|^2}{3\Mp^2}\right)^{-1}   \ .
\end{align}
Eliminating also the auxiliary vector field $\mathcal{A}_{\mu}$ one obtains, up to a total derivative,
\begin{align}\label{einstein}
\frac{1}{\sqrt{-g}}\mathcal{L} = 
\frac{1}{2}\Mp^2 R - K_{\alpha\bar{\alpha}}g^{\mu\nu}
\nabla_{\mu}z^{\alpha}\nabla_{\nu}\bar{z}^{\bar{\alpha}}  -
\Omega^4 g\partial_{\alpha}\Phi\left(T^a z\right)^{\alpha} D_a  - V \ ,
\end{align} 
with $K_{\alpha\bar{\alpha}} = \partial_{\alpha}\partial_{\bar{\alpha}}K$ 
and K\"ahler potential
\begin{align}\label{KahlerP}
K(z,\bar{z}) = -3\Mp^2 \ln \left(- \frac{1}{3\Mp^2}\Phi(z,\bar{z})\right) \ .
\end{align}
Note that the covariant derivative 
$\nabla_{\mu} = \partial_{\mu} -igA^a_{\mu}T^a$  does not contain the
auxiliary field $\mathcal{A}_{\mu}$ anymore.
For the scalar potential in the Einstein frame one obtains ($f=\delta_{ab}$),
\begin{align}\label{PotEJ}
V &= \Omega^4 \left(
G^{\alpha\bar{\alpha}} \partial_{\alpha}W \partial_{\bar{\alpha}}\overline{W}
 + \frac{1}{2} D_a^2 \right) \nonumber\\
&= e^{K/3\Mp^2} \left( K^{\alpha \bar{\alpha}}
\mathcal{D}_{\alpha} W \mathcal{D}_{\bar{\alpha}} \overline{W} 
- \frac{|W|^2}{3\Mp^2}\right)  
+ \frac{1}{2} (\Omega^2 D_a)^2 \ ,
\end{align}
where $K^{\alpha \bar\alpha}  = K^{-1}_{\alpha \bar\alpha}$ is the inverse 
K\"ahler metric, and 
$\mathcal{D}_\alpha = \partial_{\alpha} + \partial_{\alpha}K/\Mp^2$.

\subsection{Breaking superconformal symmetry}

In the Jordan frame Lagrangian (\ref{LagJ}) superconformal symmetry is
explicitly broken by the kinetic term of the gravitational field. Full
superconformal symmetry can be achieved by introducing a compensator
field $z^0$ and replacing the frame function by the $SU(1,n)$ invariant real function 
\begin{align}\label{real1}
\Xi(z^0,\bar{z}^0;z,\bar{z}) = - |z^0|^2 + |z|^2 \ .
\end{align}
The choice $z^0 = \sqrt{3}\Mp$, which corresponds to fixing a gauge
for the local conformal symmetry, then yields the frame function,
\begin{align}
\Xi(z^0,\bar{z}^0;z,\bar{z})\big|_{z^0 = \sqrt{3}\Mp} = \Phi(z,\bar{z}) \ .
\end{align}
As suggested in \cite{Ferrara:2010yw}, given a gauge singlet 
$\chi_{\alpha\beta}z^{\alpha}z^{\beta}$ with $\chi_{\alpha\beta}$
dimensionless, superconformal symmetry can be explicitly broken by 
using instead of (\ref{real1}) the real function
\begin{align}
\Xi(z^0,\bar{z}^0;z,\bar{z}) = - |z^0|^2 + |z|^2 
+\left(\chi_{\alpha\beta}\frac{z^{\alpha}z^{\beta}\bar{z}^0}{z^0} 
+ \mathrm{h.c.}\right) \ .
\end{align}
After gauge fixing one obtains the modified frame function
\begin{align}\label{framechi}
\Phi(z,\bar{z}) = -3\Mp^2 + |z|^2 + J(z) + \bar{J}(\bar{z})\ , \quad J(z)
=\chi_{\alpha\beta}z^{\alpha}z^{\beta}\ ,
\end{align}
corresponding to a Weyl rescaling between Jordan and Einstein frame with
\begin{align}
\Omega = \left(1 - \frac{1}{3\Mp^2}\left(|z|^2 + J(z) 
+ \bar{J}(\bar{z})\right)\right)^{-1/2}\ .
\end{align}
In the following analysis the symmetry breaking term $J(z)$ will play an 
important role. As we
shall see, it will turn the familiar single-field D-term inflation model
into a two-field inflation model. 

We are particularly interested in adding for a $U(1)$ gauge symmetry a
FI-term to the Lagrangian (\ref{LagJ}). Naively, this
would correspond to the substitution
$g\partial_{\alpha}\Phi Q z^{\alpha} D \rightarrow 
g\left(\partial_{\alpha}\Phi Q z^{\alpha} + \xi\right) D$, where $Q$ is the
charge generator. This, however, would introduce another explicit
breaking of superconformal symmetry, since $\xi$ is a constant of mass
dimension two. In the Lagrangian (\ref{LagJ}) superconformal symmetry 
breaking only arises from $\Xi(z^0,\bar{z}^0;z,\bar{z})$ after gauge
fixing.  This suggests to add to Eq.~(\ref{LagJ}) a term with
dimensionless constant $\hat{\xi}$,
\begin{align}
\label{eq_Lxi}
\Delta \mathcal{L}_J^{\xi} &= 
g \Xi(z^0,\bar{z}^0;z,\bar{z})\big|_{z^0 = \sqrt{3}\Mp}\hat{\xi} D  \nonumber\\
&= g \Omega^{-2} \xi D \ , 
\end{align}
where $\xi = -3 \Mp^2  \hat{\xi}$ has mass dimension two. Note that in the
Jordan frame the FI-term is field dependent.

Using Eqs.~(\ref{einstein}) and (\ref{PotEJ}) and eliminating the
auxiliary field $D$, one immediately obtains for the
D-term scalar potential in the Einstein frame,
\begin{align}\label{PotFI}
V &= \frac{g^2}{2} \left(\Omega^2\partial_{\alpha}\Phi Q z^{\alpha} + \xi\right)^2 \nonumber \\
&=\frac{g^2}{2} \left(\partial_{\alpha}K Q z^{\alpha} + \xi\right)^2 \ ,
\end{align}
which is the standard supergravity expression \cite{Wess:1992cp}. Note
that from now on we work in the Einstein frame, where we can use the standard expressions for the slow-roll parameters.

The consistency of a constant FI-term in supergravity is a subtle issue
\cite{Binetruy:2004hh,Komargodski:2009pc,Dienes:2009td,Distler:2010zg,
Arnold:2012yi}. We have in mind a field dependent, effectively constant
FI-term at the GUT scale, as it can arise in the weakly coupled heterotic
string due to the Green-Schwarz mechanism
of anomaly cancellation, where
$\xi_{\rm GS} = g_s^2 \mathrm{Tr}{Q}\Mp^2/(192\pi^2)$ \cite{Dine:1987xk}. Here $\mathrm{Tr}{Q}$
is the sum over $U(1)$ charges and $g_s$ is the string coupling, which
depends on the dilaton. Clearly, a GUT scale FI-term requires an appropriate
stabilization of the dilaton and other moduli fields (see, for example,
Refs. \cite{Binetruy:2004hh,Buchmuller:2007qf,Dundee:2010sb,Hebecker:2012aw}).
A related problem is the connection between the GUT scale and supersymmetry
breaking. A thorough discussion of these important questions goes beyond the
scope of the present paper.

\subsection{Scalar potential and F-term inflation}

Breaking superconformal symmetry by the holomorphic term $J$ in 
Eq.~(\ref{framechi}) significantly modifies the
scalar potential. From Eq.~(\ref{KahlerP}) one obtains for the frame function
$\Phi$ given in Eq.~\eqref{framechi} the K\"ahler metric
\begin{align}\label{Kmetric}
K_{\alpha\bar{\alpha}} = \Omega^2 \left(\delta_{\alpha\bar{\alpha}}
- \frac{1}{\Phi}\partial_{\alpha}\Phi \partial_{\bar{\alpha}}\Phi\right) \ .
\end{align}
One easily verifies that the inverse K\"ahler metric is given by
\begin{align}\label{Kinv}
K^{\alpha\bar{\alpha}} = \Omega^{-2} \left(\delta^{\alpha\bar{\alpha}}
+ \frac{1}{\Delta}\delta^{\alpha\bar{\beta}}\partial_{\bar{\beta}}\Phi 
\delta^{\beta\bar{\alpha}}\partial_{\beta}\Phi\right) \ ,
\end{align}
where 
\begin{align}
\Delta = \Phi - \delta^{\alpha\bar\alpha}
\partial_{\alpha}\Phi\partial_{\bar\alpha}\Phi \ .
\end{align}
Inserting Eq.~(\ref{Kinv}) into the expression (\ref{PotEJ}), one obtains for
the F-term scalar potential in the Einstein frame the compact 
expression\footnote{During the preparation of this paper, 
Ref.~\cite{Einhorn:2012ih} appeared where the same expression for the F-term 
potential has been found.}
\begin{align}\label{FtermP}
V_F = \Omega^4\left(\delta^{\alpha\bar{\alpha}}
\partial_{\alpha}W\partial_{\bar{\alpha}}\overline{W}
+ \frac{1}{\Delta}|\delta^{\alpha\bar{\alpha}}
\partial_{\alpha}W\partial_{\bar{\alpha}}\Phi - 3W|^2\right)\ .
\end{align}
Clearly, for superpotentials cubic in the fields and $J=0$, the second term 
in the bracket vanishes and one obtains the F-term potential of 
global supersymmetry up to the
rescaling factor $\Omega^2$ between Jordan and Einstein frame. 

The expression (\ref{FtermP}) holds for all superpotentials and it is
instructive to apply it to the superpotential of F-term hybrid inflation
\cite{Copeland:1994vg, Dvali:1994ms, Nakayama:2010xf, Beringer:1900zz},
\begin{align}
W = \lambda S \left( \phi_+ \phi_- -v^2\right) \ .
\end{align}
Here $\phi_\pm$ are `waterfall' fields, $v$ is a mass parameter and
$S$ contains the inflaton; the coupling $\lambda$ is chosen to be real. 

F-term hybrid inflation typically yields a
scalar spectral index which is too large compared to observation. One
may hope to improve the situation by a proper choice of the $\chi$-parameter
of the frame function
\begin{align}\label{frameinf}
\Phi = -3\Mp^3 + |S|^2 + |\phi_-|^2 + |\phi_+|^2 + 
\frac{\chi}{2}\left(S^2 + \bar{S}^2\right) \ ,
\end{align}
where we have used the same symbols for chiral superfields and their scalar 
components; the parameter $\chi$ is chosen to be real. This yields a non-minimal coupling of the inflaton field to gravity. From Eq.~(\ref{FtermP}) 
one then obtains for the scalar potential
\begin{align}\label{FPotFterm}
V_F = \Omega^4 \lambda^2\left(|S|^2\left(|\phi_+|^2 + |\phi_-|^2\right)
+ |\phi_+\phi_- - v^2|^2 
-\frac{|2 v^2 S + \chi(\phi_+ \phi_- - v^2)\bar{S}|^2}
{3\Mp^2 + \tfrac{\chi}{2}(S^2 + \bar{S}^2) 
+ {\chi^2}|S|^2}\right)  .
\end{align}
Along the expected inflationary trajectory, i.e., for $\phi_\pm = 0$, one has
\begin{align}
V_F = \Omega^4 \lambda^2 v^4 
-\frac{\Omega^4 \lambda^2 v^4 |2 S - \chi \bar{S}|^2}
{3\Mp^2 + \tfrac{\chi}{2}(S^2 + \bar{S}^2) 
+ {\chi^2}|S|^2} \ .
\end{align}
Unfortunately, this potential has a large tachyonic mass for $S$ and 
is therefore not phenomenologically viable.
Note that this could be remedied by adding an $|S|^4$ term to the frame function~\cite{Ferrara:2010yw, Lee:2010hj}. However, this introduces an additional breaking of the superconformal symmetry and we will not pursue this option here.

\subsection{D-term inflation}

Let us now consider D-term hybrid inflation. It has the attractive feature 
that in string compactifications an FI-term of GUT scale size naturally arises,
which is welcome for hybrid inflation. The superpotential reads  
\begin{align}
\label{eq_WDterm}
W = \lambda S \phi_+ \phi_- \ ,
\end{align}
and the frame function is again given by Eq.~(\ref{frameinf}).
The corresponding F-term scalar potential reads
\begin{align}\label{FPotDterm}
V_F = \Omega^4\lambda^2\left(|S|^2 (|\phi_+|^2 + |\phi_-|^2) 
+ |\phi_+ \phi_-|^2  -  \frac{\chi^2 |\phi_+|^2 |\phi_-|^2 |S|^2}
{3\Mp^2 + \frac{1}{2} \chi (S^2 + {\bar S}^2) + \chi^2 |S|^2}\right) \ .
\end{align}
This expression agrees with the potential (\ref{FPotFterm}) in the case $v=0$.
For field values below the Planck mass the potential (\ref{FPotDterm})
is well behaved. The potential vanishes identically for $\phi_\pm = 0$, which
corresponds to the inflationary trajectory.
 
The potential (\ref{FPotDterm})
is supplemented by a D-term scalar potential of a $U(1)$ gauge interaction  
under which the chiral superfields $S$ and $\phi_\pm$ have charge $0$ and 
$\pm q$, respectively. The corresponding scalar potential with nonvanishing 
FI-term is given by
\begin{align}\label{DPotDterm}
V_D = \frac{g^2}{2} \left(\Omega^2 q(|\phi_+|^2 -|\phi_-|^2)-\xi\right)^2 \ ,
\end{align}
where $g$ is the gauge coupling.
For $\phi_\pm = 0$, $V_D$ provides the vacuum energy 
$V_{0} = g^2 \xi^2/2$ which drives inflation.

The slope of the inflaton potential is generated by quantum corrections.
Along the inflationary trajectory the Weyl rescaling factor reads
\begin{align}\label{omega0}
\Omega_0 = \Omega\big|_{\phi_\pm = 0} = \left(1 - \frac{1}{3\Mp^2}
\left(|S|^2 + \frac{\chi}{2}(S^2 + \bar{S}^2)\right)\right)^{-1/2} \ .
\end{align}
From Eqs.~(\ref{PotEJ}) and (\ref{Kmetric}) one then obtains for the part 
of the Lagrangian quadratic in $\phi_\pm$,
\begin{align}
\mathcal{L}_m = \Omega_0^2 \partial_\mu \phi_\pm^* \partial^\mu \phi_\pm
-\left(\Omega_0^4\lambda^2|S|^2 \mp \Omega_0^2 q g^2 \xi\right)|\phi_\pm|^2 \ ,
\end{align} 
from which one reads off the scalar masses 
\begin{align}
 m_\pm^2 =  \Omega_0^2 \lambda^2 |S|^2 \mp q g^2 \xi \ .
\label{eq_scalar_masses}
\end{align}
For $|S|$ larger than a critical value $|S_c|$, both $\phi_+$ and $\phi_-$ 
have positive mass terms and are stabilized at zero, thus allowing 
inflation to proceed in the $S$ direction. At $|S| = |S_c|$, $m_+^2$ turns 
negative, triggering a phase transition which gives an expectation 
value to $\phi_+$ and ends inflation. The critical value $S_c$ is determined by
the relation
\begin{align}
 \Omega^2(S_c)|S_c|^2 = \frac{q g^2 \xi}{\lambda^2}\ .
\label{eq_Sc}
\end{align}

Supersymmetry is broken along the inflationary trajectory where one has 
$V > 0$. Hence, quantum corrections to the tree-level potential do not vanish
and one obtains the one-loop correction
\begin{align}
 V_{1l} = \frac{1}{64 \pi^2} \mathrm{STr} \left[ M^4 
\left( \ln\left(\frac{M^2}{Q^2}\right) - \frac{1}{2} \right) \right] \ .
\label{eq_coleman-weinberg}
\end{align}
Here $\mathrm{STr}$ denotes the supertrace running over all fields with 
$S$-dependent masses, i.e., $\phi_\pm$ and their fermionic partners.
$M$ is the corresponding mass matrix, and $Q$ is an appropriate renormalization
scale which also determines
the argument of the running gauge coupling. According to the mass sum rule, 
the Dirac fermion associated with $\phi_\pm$ has mass
\begin{align}
 m_f^2 = \lambda^2 \Omega_0^2 |S|^2 \ .
\label{eq_fermion_mass}
\end{align}
Inserting Eqs.~\eqref{eq_scalar_masses} and \eqref{eq_fermion_mass} into 
Eq.~\eqref{eq_coleman-weinberg} and choosing the renormalization scale 
$Q^2 = g^2 q \xi$, one obtains for the one-loop potential, 
\begin{align}\label{eq_1-loop}
V_{1l} &= \frac{g^4 q^2\xi^2}{32\pi^2}\left((x-1)^2\ln(x-1) + (x+1)^2 \ln(x+1) 
- 2 x^2 \ln x  -1 \right) \nonumber \\
&= \frac{g^4 q^2\xi^2}{16\pi^2} \left(1 + \ln x 
+ {\cal O}\left(\frac{1}{x}\right) \right) \ , 
\end{align}
where
\begin{align}
x = \frac{\lambda^2 \Omega_0^2 |S|^2}{q g^2 \xi} 
= \frac{\Omega_0^2(S) |S|^2}{\Omega_0^2(S_c) |S_c|^2}\ .
\end{align}
The total potential is given by (cf.~Eqs.~(\ref{FPotDterm}), (\ref{eq_1-loop}))
\begin{align}
V &= (V_F + V_D + V_{1l})\big|_{\phi_\pm = 0} \nonumber\\
&= \frac{g^2}{2}\xi^2 \left(1 + \frac{g^2 q^2}{8\pi^2} 
\left(1 + \ln x + {\cal O}\left(\frac{1}{x}\right)\right)\right)\ .
\end{align}
Note that on the inflationary trajectory one has $|S| > |S_c|$ and $x > 1$.

In this section, we calculated the one-loop correction to the scalar potential in the Einstein frame for the Minkowski metric, $g_{\mu \nu} = \eta_{\mu \nu}$. Note that a calculation in the Jordan frame would have led to the same result. One then starts from Eqs.~\eqref{LagJ}, \eqref{PotJ}, \eqref{eq_Lxi} and \eqref{eq_WDterm}. The quantum correction to the potential depends on the background metric, which in the Jordan frame is given by $g_{J  \mu \nu} = \Omega_0^2 \; \eta_{\mu \nu}$, cf. Eqs. \eqref{eq_gJE} and \eqref{omega0}. One easily verifies explicitly that the corresponding scalar masses $m^J_\pm$ are identical with the masses given in Eq.~\eqref{eq_scalar_masses}. This leads to the one-loop correction for the scalar potential $\sqrt{- g_J} \, V^J_{1l} = V_{1l}$ (cf. Eq.~\eqref{eq_coleman-weinberg}), i.e.\ $V_{1l}^J = \Omega_0^{-4} V^J_{1l}$. Transforming back to the Einstein frame, one obtains $V_{1l}$ as one-loop correction to the scalar potential, in agreement with the calculation performed directly in the Einstein frame.


\section{Single-field inflation \label{sec_single_field}}
\subsection{Slow-roll equation of motion \label{subsec_slowrolleom}}

We are now ready to tackle the slow-roll equations of motion for the field $S$.
Note that the inflaton field $S$ is not canonically normalized, which leads to
a modification of the standard slow-roll equations.

Expressing the Lagrangian for the field $S$ in terms of real and imaginary
components, $S = (\sigma + i \tau)/\sqrt{2}$,
\begin{equation}
\frac{1}{\sqrt{-g}}\mathcal{L} = \frac{1}{2} K_{S \bar{S}}(\sigma,\tau) 
(\partial_{\mu} \sigma \partial^{\mu} \sigma + 
\partial_{\mu} \tau \partial^{\mu} \tau) - V(\sigma, \tau) \ ,
\label{eq_lagrange}
\end{equation}
where $g_{\mu \nu}$ now denotes the FRW metric,
one obtains the slow-roll equations for the homogeneous fields $\sigma$ 
and $\tau$,
\begin{equation}
 3 K_{S \bar{S}} H \dot \sigma = - \frac{dV_{1l}}{d \sigma}\ ,  \quad   
3 K_{S \bar{S}} H \dot \tau = - \frac{dV_{1l}}{d \tau}\ ,
\label{eq_double_sr}
\end{equation}
where now we have set $\Mp = 1$ for convenience. These equations can be 
written
as the standard slow roll equations for an effective potential defined by
\begin{equation}
\frac{dV_{eff}}{d \sigma} = \frac{1}{K_{S \bar{S}}}\frac{dV_{1l}}{d \sigma}\ , 
\quad  
\frac{dV_{eff}}{d \tau} = \frac{1}{K_{S \bar{S}}} \frac{dV_{1l}}{d \tau}\ .
\label{eq:dVeffdphi}
\end{equation}
Calculating the second derivatives of the potential $V_{eff}$ with respect 
to $\sigma$ and $\tau$, one finds that for $\chi < 0$, the trajectory 
$\sigma \neq 0, \tau = 0$ yields a viable inflationary trajectory along which
$d^2 V_{eff}/d \tau^2$ is positive. Hence this trajectory is an attractor for 
a sufficiently long phase of inflation before the onset of the final 50 
e-folds. 
For $\chi >0$, the situation is reversed and an equivalent inflationary
trajectory corresponds to $\sigma = 0, \tau \neq 0$. For $\chi = 0$, the 
Lagrangian is independent of the phase of $S$ and the inflaton can be 
identified as the absolute value of $S$. In the following we choose 
$\chi \leq 0$.

In this section we will restrict ourselves to the standard case of `one-field' inflation
described above, postponing the discussion of possible two-field inflation 
to Section~\ref{sec_two-field}. Inserting the K\"ahler metric
\begin{align}
K_{S \bar{S}}\big|_{\phi_\pm,\tau =0} = 
\frac{1}{1-\frac{1}{6}(1+\chi)\sigma^2}\left(1 -
\frac{(1+\chi)^2\sigma^2}{6\left(1-\frac{1}{6}(1+\chi)\sigma^2\right)}\right)
\end{align}  
and the one-loop potential (\ref{eq_1-loop}) into the slow-roll equation
(\ref{eq_double_sr}), one obtains after integrating from  $\sigma_*$ to 
$\sigma_f$, 
\begin{equation}
 3 \ln \left( \frac{1 - \frac{1}{6}(1 + \chi) \sigma_*^2}{1 - \frac{1}{6}(1 + \chi) \sigma_f^2}\right) - \frac{1}{2} \chi \left(- \sigma_*^2 + \sigma_f^2\right) \simeq - \frac{g^2 q^2}{4 \pi^2} N_*  \ .
\label{eq_sigmae}
\end{equation}
Here $\sigma_f$ denotes the value of $\sigma$ at the end of inflation and 
$\sigma_*$ is the value of $\sigma$ $N_*$ e-folds earlier. 
Inflation ends when either $m^2_+$ turns negative ($\sigma_f = \sigma_c$) 
or when the slow-roll conditions are violated ($\sigma_f = \sigma_{\eta}$). 
From Eq.~\eqref{eq_Sc} and Eq.~\eqref{eq_epsilon_eta} with $|\eta| = 1$, 
one finds
\begin{equation}
\sigma_c^2 = \frac{6 g^2 q \xi}{3 \lambda^2 + (1 + \chi) g^2 q \xi}\ ,  
\qquad \sigma_{\eta}^2 \approx \frac{g^2 q^2}{4 \pi^2} \ .
\label{eq_sigma_f}
\end{equation}

For small field values, satisfying $|1+\chi|\sigma_*^2/6 \ll 1$, 
Eq.~(\ref{eq_sigmae}) can be solved analytically, leading to 
\begin{equation}\label{smallfield}
 \sigma_*^2 \simeq \sigma_f^2 + \frac{g^2 q^2}{2 \pi^2} N_* \ .
\end{equation}
However, for most of parameter space this is a bad approximation, and
one has to solve Eq.~(\ref{eq_sigmae}) numerically.

\subsection{Slow-roll parameters \label{sec_slow-roll-parameters}}

In order to calculate the spectral index and other observables, we need to 
evaluate the slow-roll parameters
\begin{equation}
 \epsilon = \frac{1}{2}\left( \frac{V'(\hat\sigma)}{V} \right)^2\ , \qquad 
 \eta = \frac{V''(\hat\sigma)}{V} \ .
\end{equation}
Here $\hat{\sigma}$ is the canonically normalized inflaton field which is
determined by (cf.~Eq.~(\ref{eq_lagrange}))
\begin{equation}
 \frac{d\sigma}{d \hat{\sigma}} = \frac{1}{\sqrt{K_{S \bar{S}}}} \ .
\end{equation}
On the inflationary trajectory
the derivatives of the scalar potential with respect to $\hat \sigma$  
can be written as
\begin{equation}
 \begin{split}
  V'(\hat \sigma) &= \frac{dV_{1l}}{d \hat\sigma} 
= \frac{dV_{1l}}{d\sigma} \frac{d\sigma}{d \hat\sigma} \ , \\
V''(\hat \sigma) &=  \frac{d^2 V_{1l}}{d \hat \sigma^2} 
= \frac{d \sigma}{d \hat \sigma} \frac{d}{d \sigma} 
\left( \frac{dV_{1l}}{d\hat\sigma} \right) \ ,
 \end{split}
\label{eq_dV}
\end{equation}
from which one obtains the slow-roll parameters
\begin{equation}
\begin{split}
\epsilon &\simeq 2 \left( \frac{g^2 q^2}{8 \pi^2}\right)^2 \frac{1}{\sigma^2} 
\frac{1}{1 + \frac{1}{6} \chi (1 + \chi) \sigma^2} \ , \\
\eta &\simeq - \frac{g^2 q^2}{4 \pi^2} \frac{1}{\sigma^2} 
\frac{(1 - \frac{1}{6} (1 + \chi) \sigma^2)(1 + \frac{1}{3} \chi (1 + \chi) 
\sigma^2)}{(1 + \frac{1}{6} \chi (1 + \chi) \sigma^2)^2} \ .
 \end{split}
\label{eq_epsilon_eta}
\end{equation}
Note that for $\chi = -1$, one obtains the results for D-term inflation
in global supersymmetry.

\subsection{Results and discussion \label{sec_results}}
\begin{figure}[t]
\center
\includegraphics[width = 0.6\textwidth]{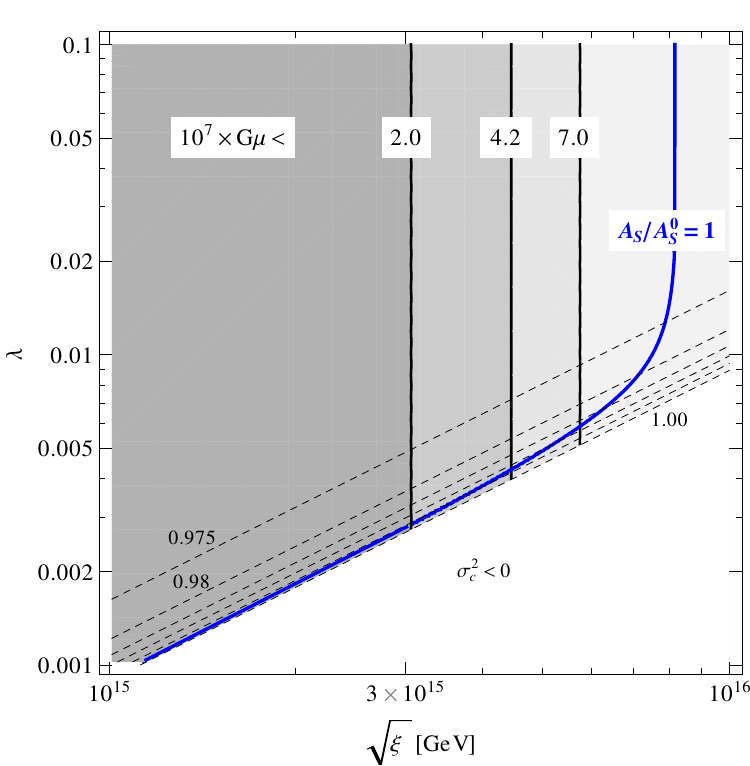}
\caption{Normalization condition and cosmic string bound for $\chi = -15$, $q = 2$, $g^2 = 1/2$ 
and $N_* = 50$. The blue line shows the relationship between $\xi$ and 
$\lambda$ imposed by the correct normalization of the amplitude of the 
primordial fluctuations. The black lines denote the cosmic string bound 
for $G \mu \times 10^7 < 2, 4.2$ and $7$, respectively; the darker shaded regions 
on the left are in agreement with the constraint. The dashed lines show 
contours of constant scalar spectral index. The white region to the bottom right must be excluded since there is no positive solution to $m^2_{+}(\sigma_c) = 0$.} 
\label{fig_cobe_strings}
\end{figure}

\vspace{0.3cm}
\noindent\textbf{Normalization of the scalar power spectrum and cosmic 
strings}
\vspace{0.3cm}

\noindent
The normalization condition for the amplitude of the primordial power spectrum 
and the cosmic string bound represent observational constraints which have to 
be fulfilled by a viable model. Allowing for a cosmic string contribution to 
the power spectrum of the primordial fluctuations implies extending the usual 
six parameter $\Lambda$CDM fit to the CMB data by an additional parameter 
which accounts for the cosmic string contribution. Detailed analyses for
Nambu-Goto strings and Abelian Higgs (AH) cosmic strings have been carried out
by several groups \cite{Battye:2010xz,Dunkley:2010ge,Urrestilla:2011gr,
Dvorkin:2011aj}. In the waterfall transition ending D-term hybrid inflation, a local $U(1)$ symmetry is broken and
AH cosmic strings may be formed. In the following discussion we shall 
therefore use the results of the recent analysis in 
Ref.~\cite{Urrestilla:2011gr} which is based on the field theoretical 
simulation of cosmic strings in Ref.~\cite{Bevis:2006mj}.
 
The analysis in Ref.~\cite{Urrestilla:2011gr} yields an upper bound on a 
cosmic string contribution of about $5\%$ and a best-fit value 
for the amplitude of the scalar contribution to the primordial fluctuations,
\begin{equation}
 A^0_s = (2.15^{+0.07}_{-0.06}) \times 10^{-9} \ ,
\label{eq_As0}
\end{equation}
where a $1 \sigma$ error has been given. 
Comparing this value with the expression calculated in our model,
\begin{equation}
A_{\text{s}}  = \frac{1}{12 \pi^2} \frac{V^{3}}{V'^2}\Big|_{\sigma =\sigma^*} \label{eq_As1} \,, 
\end{equation}
which, using Eq.~\eqref{eq_1-loop}, can be simplified to
\begin{equation}
A_{\text{s}}   \simeq \frac{2 \pi^2}{3} \frac{\xi^2 \sigma_*^2}{g^2 q^4} 
\left[1 + \frac{1}{6} \chi (1 + \chi) \sigma_*^2\right] \ ,
\label{eq_As2}
\end{equation}
one obtains a relation between $\xi$ and $\lambda$ for given values of 
$\chi$, $q$ and $g$. 

As an example, we choose $q=2$ and $g=1/\sqrt{2}$ in the following, 
which is motivated by identifying the spontaneously broken $U(1)$ symmetry 
with $U(1)_{B-L}$ (c.f.~\cite{Buchmuller:2012wn}).
For $\chi=-15$, the implied relation between  $\xi$ 
and $\lambda$ is represented by the blue line in Fig.~\ref{fig_cobe_strings}. 

A fit to the CMB data assuming scalar perturbations and AH cosmic strings
yields an upper bound on $G \mu$, where $G= (8 \pi \Mp^2)^{-1}$ is Newton's 
constant and
\begin{align}
\mu = 2 \pi \langle \phi_+ \rangle^2 B(\beta)
\end{align}
denotes the string tension. Here $B(\beta)$, with $\beta$ parametrizing the ration of the U(1) vector boson and the inflaton masses in the true vacuum,  $\beta = 2 m_V^2/m_{\sigma}^2 $, gives the deviation from the 
Bogomolnyi bound, with $B(2) = 1$. 
Inserting the vacuum expectation value of the waterfall field, $\langle \phi_+ \rangle = (\xi/q)^{1/2}$, as well as $m^2_V = m_\sigma^2 = 2 g^2 q \xi$, one obtains
\begin{equation}\label{stringbound}
 G \mu = 5.3 \times 10^{-7} \, \frac{2}{q} \, 
\frac{\xi}{\left(5 \times 10^{15}~\mathrm{GeV}\right)^2} \ .
\end{equation}
This is to be compared with the $2\sigma$ upper bound found in the analysis of Ref.~\cite{Urrestilla:2011gr},
\begin{equation}
 G \mu < 4.2 \times 10^{-7} \ .
\label{eq:Gmubound}
\end{equation}
The solid black lines  in Fig.~\ref{fig_cobe_strings} correspond to the string 
tensions $G \mu = (2,4.2,7) \times 10^{-7}$. The brighter region on the 
right of a given line is excluded, whereas the darker region on 
the left is in agreement with the respective bound.

The upper bounds on the string tension have a considerable theoretical
uncertainty. For instance, the upper bounds for Nambu-Goto strings are 
more restrictive than the ones for AH cosmic strings by about a factor 
of three \cite{Battye:2010xz}. This can be traced back to decay channels
into massive radiation for AH cosmic strings \cite{Hindmarsh:2008dw}.
Note also, that all simulations have been done for a bosonic Abelian
Higgs model, whereas in D-term inflation one is considering a supersymmetric
theory. Additional fermionic decay channels may further relax the cosmic
string bound by a factor ${\cal{O}}(1)$. Last but not least, one has to worry
about initial conditions. Clearly, strings cannot form until the causal
horizon is larger than their characteristic width \cite{Bevis:2006mj}, 
and one should remember that tachyonic preheating proceeds very fast. In fact,
the expectation value $\langle|\phi_+|^2\rangle$ of the waterfall field 
grows with time faster than exponentially \cite{Asaka:2001ez}. 

\vspace{0.6cm}
\noindent \textbf{Spectral index}
\vspace{0.3cm}

\begin{figure}
\centering
\subfigure[]{
\includegraphics[width = 0.45\textwidth]{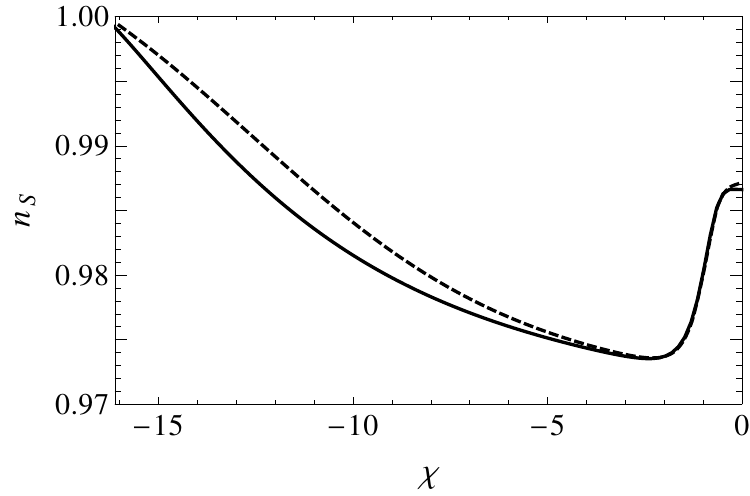}
\label{fig_ns}} \hspace{0.1cm}
\subfigure[]{
\includegraphics[width = 0.45\textwidth]{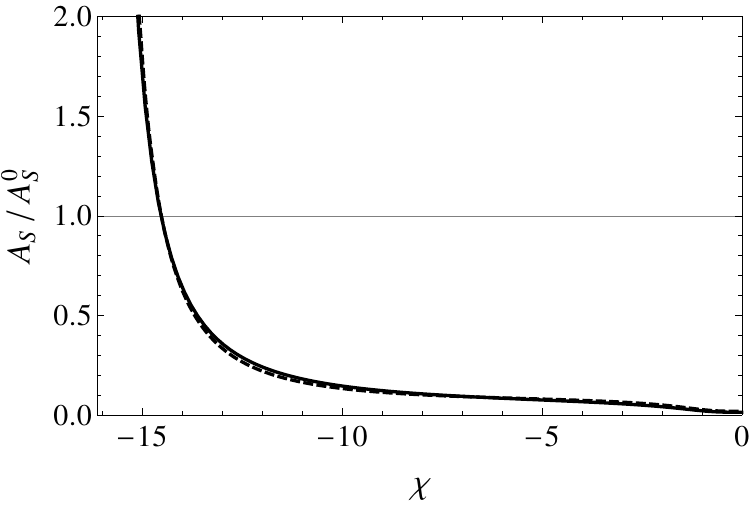}
\label{fig_P}} 
\caption{Spectral index and total amplitude for $\sqrt{\xi} = 4.3 \times 10^{15}$~GeV, $\lambda = 4 \times 10^{-3}$, $q = 2$, $g^2 = 1/2$, $N_* = 50$. The solid lines show the numerical results, the dashed lines the analytical ones. The values of $\xi$ and $\lambda$ are chosen such as to be compatible with the cosmic string bound as well as the normalization constraint for $\chi = -15$ 
(c.f.~Fig.~\ref{fig_cobe_strings}).} 
\label{fig_ns_P}
\end{figure}

\noindent With the slow-roll parameters from Eqs.~\eqref{eq_epsilon_eta} and the value of $\sigma$ $N_*$ e-folds before the end of inflation, cf.~Eq.~\eqref{eq_sigmae}, at hand, we can now easily calculate the spectral index,
\begin{equation}
 n_s = \left(1 - 6 \epsilon + 2 \eta\right)\big|_{\sigma = \sigma_*}\ .
\end{equation}
Fig.~\ref{fig_ns} shows the resulting $\chi$ dependence for a ($\xi, \lambda$) pair compatible with the cosmic string bound and the normalization condition 
at $\chi = -15$ (cf.~Fig.~\ref{fig_cobe_strings}). For reference, Fig.~\ref{fig_P} shows the corresponding $\chi$-dependence of the total amplitude. Both curves are shown over the entire range of allowed $\chi$-values for this choice of $\xi$ and $\lambda$, which is bounded from below by the condition that $\sigma_c^2$ in Eq.~\eqref{eq_sigma_f} is positive.

The dashed lines show the results obtained using the analytical formulas~\eqref{eq_epsilon_eta} and \eqref{eq_As1} with $\sigma_*$ determined by Eq.~\eqref{eq_sigmae}, the solid lines show the full numerical results. 
The deviation visible in Fig.~\ref{fig_ns} is due to the approximation of the one-loop potential, which enters in the derivation of Eq.~\eqref{eq_sigmae} and in the expressions for the slow-roll parameters $\epsilon$ and $\eta$. To obtain the numerical result, we do not use this approximation, but proceed with the full expression given in the first line of Eq.~\eqref{eq_1-loop}.
Note, however, that these corrections only influence the result for the spectral index at the per mille level, proving that the analytical results obtained above do indeed give a good description of the quantitative results.

Throughout the parameter region compatible with the normalization condition and the cosmic string bound, the spectral index is rather high, $n_s \simeq 0.99 - 1.0$. However, 
taking into account a contribution of cosmic strings close to the current bound significantly modifies the best-fit value of $n_s$ to the CMB data compared to the standard six parameter $\Lambda$CDM fit.
In Ref.~\cite{Urrestilla:2011gr}, the spectral index matching the amplitude given in Eq.~\eqref{eq_As0} and a cosmic string contribution of about $2 \%$ is found to be
\begin{align}
n_s = 0.969 \pm 0.013 \ . 
\end{align}
The obtained values for the spectral index are thus compatible with
current observational data at about the $2\sigma$ level.

The qualitative behaviour of the relation between the coupling 
$\lambda$ and the inflationary energy scale $\sqrt{\xi}$, displayed in
Fig.~\ref{fig_cobe_strings}, can be easily understood. In the case of
small coupling, $\lambda \lesssim 0.01$, one has 
$\sigma_*^2 \simeq \sigma_c^2$ (cf.~Eq.(\ref{eq_sigma_f})). The correct 
fluctuation amplitude is then 
obtained for small values of $\sqrt{\xi}$ and the cosmic string
bound can be satisfied. However, the field value $\sigma_*$ is large,
and one therefore obtains a large spectral index, $n_s \simeq 1$.

On the other hand, for large couplings $\lambda$, one has $\sigma_f^2 \ll 1$. 
For large values of $(-\chi)$,  Eq.~(\ref{eq_sigmae}) then
implies for the field value $\sigma_*$ at $N_*$ e-folds,
\begin{align}
\sigma_*^2 \simeq - \frac{g^2q^2 N_*}{2\pi^2\chi} \ .
\end{align}
Interestingly, the amplitude of scalar fluctuations is then only determined
by the energy density during inflation, $V_0 = g^2\xi^2/2$ (cf.~Eq.~(\ref{eq_As2})),
\begin{align}
A_s \simeq \frac{V_0}{18\pi^2} N_*^2 \ .
\end{align} 
For the spectral index one finds\footnote{Note the difference to D-term
inflation in global supersymmetry, where one has 
$n_s \simeq 1 - \frac{1}{N_*} \simeq 0.98$, see Ref.~\cite{Kallosh:2003ux}.}
\begin{align}
n_s  \simeq (1-2\eta)\big|_{\sigma_*} \simeq 1 - \frac{2}{N_*} \simeq 0.96\ .
\end{align}

Contrary to the amplitude of scalar fluctuations, the string tension
additionally depends on the coupling strength $gq$ (cf.~Eq.~(\ref{stringbound})),
\begin{align}
 G \mu = 5.3 \times 10^{-7} \, \frac{2\sqrt{2}}{gq} \, 
\frac{V_0^{1/2}}{\left(5 \times 10^{15}~\mathrm{GeV}\right)^2} \ .
\end{align}
Hence, for large values of $(-\chi)$ and $\lambda$, it is always possible
to satisfy the cosmic string bound by increasing $gq$ while at the
same time keeping $n_s$ small. This is in contrast
to the case where $|1+\chi|\sigma_*^2/6 \ll 1$ and $\sigma_c^2 \ll
\sigma_*^2$, with $\sigma_*$ given by 
Eq.~(\ref{smallfield}). In this case the amplitude is given by
$A_s \simeq 2\epsilon N_*/(3g^2q^2)$ whose value also fixes the string
tension. However, increasing $gq$ one moves to a regime of
strong coupling and the theoretical consistency of the model becomes 
questionable.

For the other CMB observables, i.e., the tilt of the spectral index $\alpha_s$ and the tensor to scalar ration $r$, we find small values, well within the experimental bounds~\cite{Komatsu:2010fb}. For instance, for the parameter point discussed above, $\sqrt{\xi} = 4.3 \times 10^{15}$~GeV, $\lambda = 4 \times 10^{-3}$, $q = 2$, $g^2 = 1/2$, $\chi = -15$ and $N_* = 50$, one obtains
\begin{equation}
 \begin{split}
  \alpha_s &= 16 \epsilon \eta - 24 \epsilon^2 - 2 \frac{V' V'''}{V^2} \, \big|_{\sigma = \sigma_*} = - 1.7 \times 10^{-4} \,, \\
 r &= 16 \epsilon \, \big|_{\sigma = \sigma_*} = 6.4 \times 10^{-6} \ .
 \end{split}
\end{equation}

In conclusion, Fig.~\ref{fig_cobe_strings} shows that there is a considerable region in parameter space, which is compatible with the normalization condition as well as cosmic string bounds. However, for generic gauge coupling strengths
$gq$, this implies a rather large value for the spectral index. Vice
versa, in the region of parameter space which yields a spectral index close to the best fit value $n_s \simeq 0.97$, we find a cosmic string tension exceeding the cosmic string bound. 
In the viable region of parameter space in between these two limiting
cases, we thus find a high contribution of cosmic strings close the
current bounds as well as a value for the spectral index which is
slightly larger than the current best-fit value. 
Clearly, upcoming experiments will provide further stringent tests of
superconformal D-term hybrid inflation.

It is worth stressing that the discussed parameter region allows for large values of the gauge coupling constant $g$, compatible with grand unification.
In this respect, the model presented here differs significantly from D-term inflation with  canonical K\"ahler potential. In the latter case, the masses entering the one-loop potential carry $\exp(|S|^2)$ factors, leading to problems for the super-Planckian values of $|S|$ typically obtained in D-term inflation. Avoiding this forces the gauge coupling $g$ to be small, $g \lesssim 2 \times 10^{-2}$, as found in Ref.~\cite{Rocher:2004et}.

\section{Two-field inflation \label{sec_two-field}}

\subsection{Two-field versus single-field inflation \label{subsec_twovssingle}}
In the previous section, we focused on the situation where one of the two real degrees of freedom of the complex scalar field $S$ plays the role of the inflaton, whereas the value of the other degree of freedom is fixed at zero. This is the case if either the second degree of freedom has a mass of order of the Hubble scale or if inflation before the onset of the final $50$ e-folds lasted sufficiently long, so that the inflationary trajectory in the direction of the smallest curvature has become an attractor. Here, with the mass difference between $\sigma$ and $\tau$ governed by the symmetry breaking parameter $\chi$, typically both masses are below the Hubble scale, resulting in a two-field inflation model. This section is hence dedicated to investigating alternative possible trajectories in ($\sigma, \tau$) field space.

\begin{figure}[t]
\center
 \includegraphics[width = 0.99\textwidth]{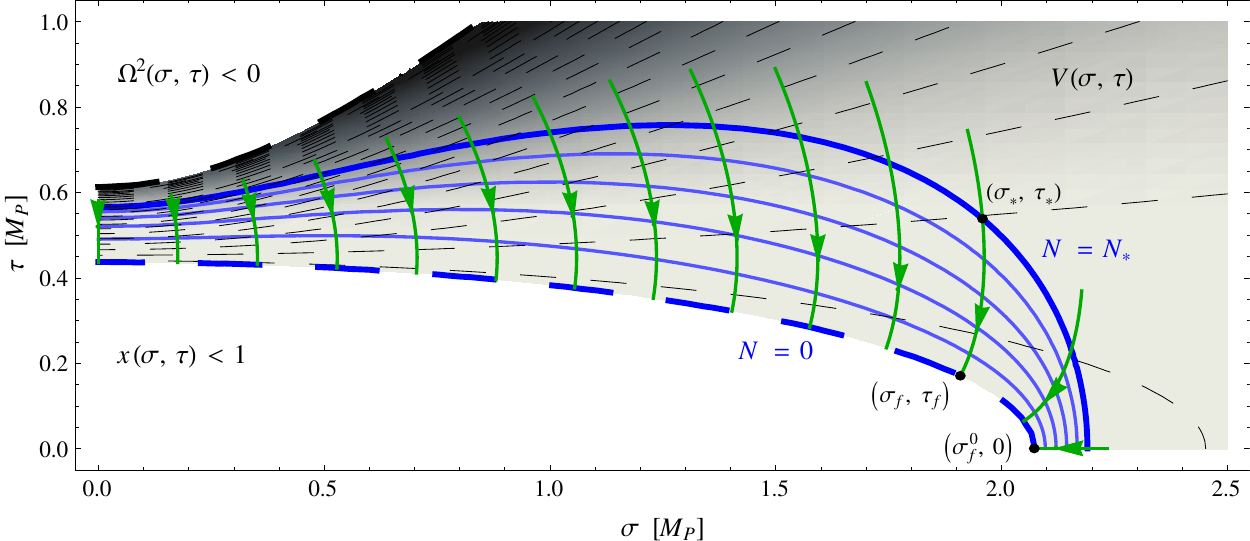}
\caption{Inflationary trajectories in ($\sigma, \tau$) field space for $\chi = -15$, $\lambda = 4 \times 10^{-3}$, $\sqrt{\xi} = 4.3 \times 10^{15}$~GeV, $g^2 = 1/2$ and $q = 2$. Contour lines of the scalar potential are denoted by dashed lines. The dashed blue line marks the $m_+ = 0$ condition, the green solid lines show several examples of inflationary trajectories. The blue lines show contours of the number of e-folds $N$, from $N= 0$ to $N = N_* = 50$. The single field case discussed in Section~\ref{sec_single_field} corresponds to the trajectory coinciding with the $\sigma$-axis.}
\label{fig_trajectories}
\end{figure}

In single-field hybrid inflation, inflation ends at the critical value of the inflaton field, $\sigma_f$, determined by the zero point of the mass of the waterfall field, ${m_+(\sigma_f)=0}$.\footnote{Here and in the following, we assume that the slow-roll conditions hold until the inflaton field reaches its critical value.} The starting point $\sigma_*$ of the inflationary trajectory is determined by solving the slow-roll equation.
In two-field inflation, the condition $m_+(\sigma_f, \tau_f) = 0$ defines a line in ($\sigma, \, \tau$) field space. From each point  on this line ($\sigma_f, \, \tau_f(\sigma_f)$), a classical inflationary trajectory can be uniquely determined by solving the set of slow-roll equations~\eqref{eq_double_sr}. The resulting trajectory ends at ($\sigma_*(\sigma_f), \tau_*(\sigma_f)$). The single-field case discussed in Section~\ref{sec_single_field} is reproduced for $(\sigma_f, \tau_f) = (\sigma_f^0, 0)$, where $\sigma_f^0$ is given by Eq.~\eqref{eq_sigma_f}. 
Hence in two-field inflation, as opposed to single-field inflation, the inflationary predictions are not uniquely determined by the parameters of the Lagrangian, but depend on an additional parameter which labels the various possible trajectories. In the notation above, this additional parameter is $\sigma_f$. This is illustrated in Fig.~\ref{fig_trajectories}. 

A generalization of the usual single-field formulas for the amplitude of the scalar fluctuations and the spectral index to the case of multi-field inflation with a non-trivial metric in field space can be found in Ref.~\cite{Sasaki:1995aw}. Starting from the action
\begin{equation}
 S = \int   d^4 x \sqrt{-g} \left[ \frac{1}{2} h_{ab} g^{\mu \nu} \partial_{\mu} \phi^a \partial_{\nu} \phi^b - V(\phi) \right]  \,,
\end{equation}
with $g_{\mu \nu}$ denoting the spacetime metric, $h_{ab}$ the metric on the real scalar field space and $\phi^a$ the real scalar fields of the theory, 
 the slow-roll conditions read
\begin{equation}
 (\partial^a V) (\partial_a V) \ll V^2 \quad \text{and} \quad \sqrt{ (\nabla^b \partial^a V ) (\nabla_b \partial_a V)} \ll V \,.
\end{equation}
Here the usual partial derivatives and covariant derivatives in scalar field space are denoted by $\partial_a = \partial / \partial \phi^a$ and $\nabla_a X^b= \partial_a X^b + \Gamma^b_{\;\;ca} X^c$.
As usual, the metric $h_{ab}$ can be used to raise or lower indices. 
For inflationary trajectories satisfying these conditions, 
the authors of Ref.~\cite{Sasaki:1995aw} obtain the following expressions for the amplitude of the primordial power spectrum and the spectral index:
\begin{equation}
 \begin{split}
  P_{s} &=  \left( \frac{H^2}{2 \pi} \right)^2 h^{ab} (\partial_a N) (\partial_b N) \,, \\
n_{s} -1 &= \frac{\left[2 \, \nabla_b \partial^a \ln V + \left(\frac{2}{3} R^a_{\;\; cbd} - h^a_{\;\; b} h_{cd} \right) (\partial^c \ln V) (\partial^d \ln V) \right] (\partial_a N) (\partial^b N)}{(\partial_e N) (\partial^e N)} \,,
 \end{split}
\label{eq_two-field-pred}
\end{equation}
with $N$ denoting the number of e-folds, $h^{ab}$ the inverse metric, $h^a_{\;\; b} = \delta^a_b$ and $R^a_{\;\; bcd}$ the scalar field space curvature tensor, $R^a_{\;\; bcd} = \partial_c \Gamma^a_{\;\; bd} - \partial_d \Gamma^a_{\;\; bc} + \Gamma^a_{\;\;ce} \Gamma^e_{\;\; db} - \Gamma^a_{\;\; de}  \Gamma^e_{\;\; cb}$ with the Christoffel symbols $\Gamma^a_{\;\; bc} = \frac{1}{2} h^{ad} (\partial_c h_{db} + \partial_b h_{dc} - \partial_d h_{bc})$.

The number of e-folds $N$ as a function of the scalar fields $\phi^a$ is determined by integrating along all possible classical trajectories. Each point in field space lies on exactly one classical trajectory. Integrating along this trajectory yields the value of $N$ at this point in field space, which is illustrated by the solid blue contour lines in Fig.~\ref{fig_trajectories}.

\begin{figure}
\centering
\subfigure[]{
\includegraphics[width = 0.45\textwidth]{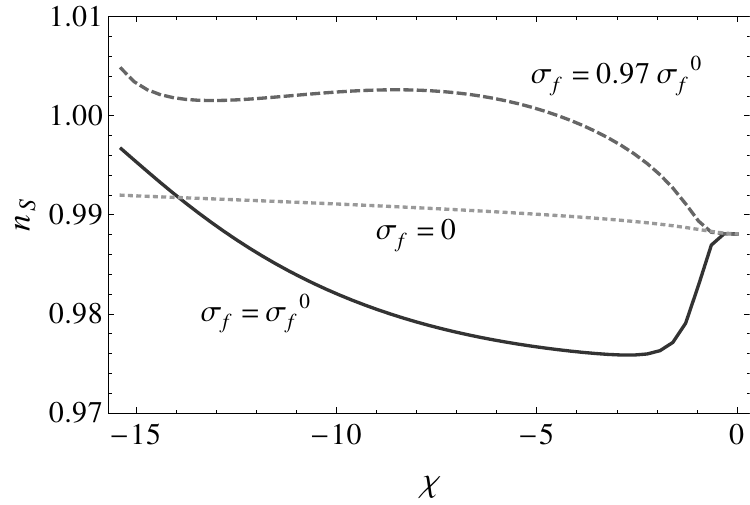}
\label{fig_ns2}} \hspace{0.1cm}
\subfigure[]{
\includegraphics[width = 0.45\textwidth]{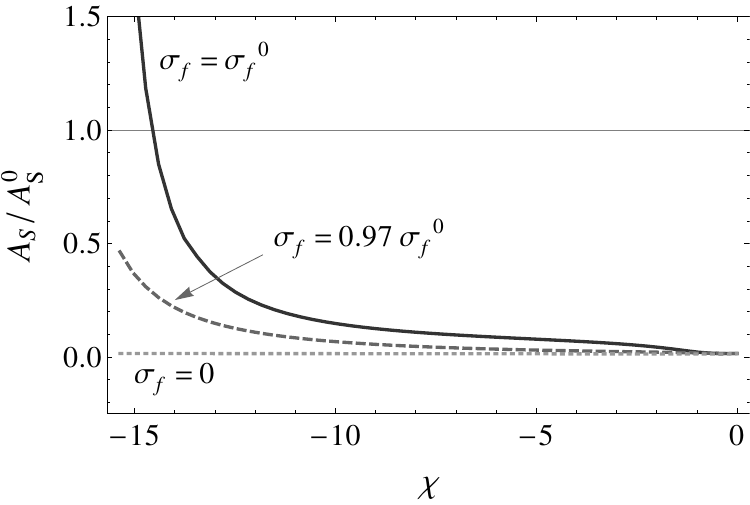}
\label{fig_P2}} 
\caption{Spectral index and total amplitude resulting from different inflationary trajectories for the same values of model parameters as in the single-field case depicted in Fig.~\ref{fig_ns_P}, i.e.\ $\sqrt{\xi} = 4.3 \times 10^{15}$~GeV, $\lambda = 4 \times 10^{-3}$, $q = 2$, $g^2 = 1/2$, $N_* = 50$.}
\label{fig_ns_P_2field}
\end{figure}

\subsection{Two-field results}


Fig.~\ref{fig_ns_P_2field} shows the spectral index and the
amplitude of the scalar power spectrum corresponding to different inflationary trajectories.
The solid lines represent the results for the trajectory along the
$\sigma$-axis, i.e.\ for $\sigma_f = \sigma_f^0$, hence reproducing the single-field
results depicted in Fig.~\ref{fig_ns_P}.
The dotted lines correspond to the other extremal case in which the inflationary trajectory
runs along the $\tau$-axis, i.e., in which $\sigma_f = 0$.
Finally, the dashed lines show the results for an intermediate trajectory with non-trivial
evolution in both $\sigma$- and $\tau$-direction.


As illustrated in Fig.~\ref{fig_P2}, the amplitude of the scalar power spectrum becomes smaller
the more the inflationary trajectory deviates from the $\sigma$-axis.
Naively, one might expect a different behaviour, since the gradient of $N$ becomes large for inflationary trajectories along the $\sigma$- as well as the $\tau$-axis, cf.\ Fig.~\ref{fig_trajectories}.
But for negative $\chi$ the entries of the inverse K\"ahler metric,
$h^{\sigma\sigma} = h^{\tau\tau} = K^{S\bar{S}}$, become increasingly smaller
the further one moves along the $N = N_*$ contour away from the $\sigma$-axis.
As it turns out, this decrease in $K^{S\bar{S}}$ dominates over 
the change in
the gradient of the number of e-folds, so that the amplitude ends up going down
as soon as one chooses an inflationary trajectory other than the one discussed in
Section~\ref{sec_single_field}.
In order to understand the decrease in the amplitude more intuitively, it is
useful to consult the single-field expression for $A_s$ in Eq.~\eqref{eq_As2}.
Interpreting $V'$ appearing in this expression as the derivative of the scalar potential
along the respective inflationary trajectory, the single-field expression for
$A_s$ may serve as a lowest-order approximation of the full multi-field expression
in Eq.~\eqref{eq_two-field-pred}.
From Eq.~\eqref{eq_As2} it is then apparent that a steeper potential, i.e., a larger
$V'$, entails a smaller amplitude.
Since for negative $\chi$ the scalar potential indeed becomes steeper
the further one moves along the $N = N_*$ contour towards the $\tau$-axis, this
explains our observation in Fig.~\ref{fig_P2}.


The behaviour of the scalar spectral index $n_s$ is more complicated.
We find that typically the minimal value of $n_s$ as a function of $\chi$ is enhanced when considering trajectories involving a motion in $\tau$-direction.
In the limit $\chi \rightarrow 0$, the three curves for the scalar spectral index
as well as the amplitude in Fig.~\ref{fig_ns_P_2field} respectively converge to common values.
This reflects the fact that for $\chi = 0$ the phase of the complex inflaton field $S$
turns unphysical, rendering all possible trajectories equivalent to each other.


\begin{figure}
\center
\includegraphics[width = 0.9\textwidth]{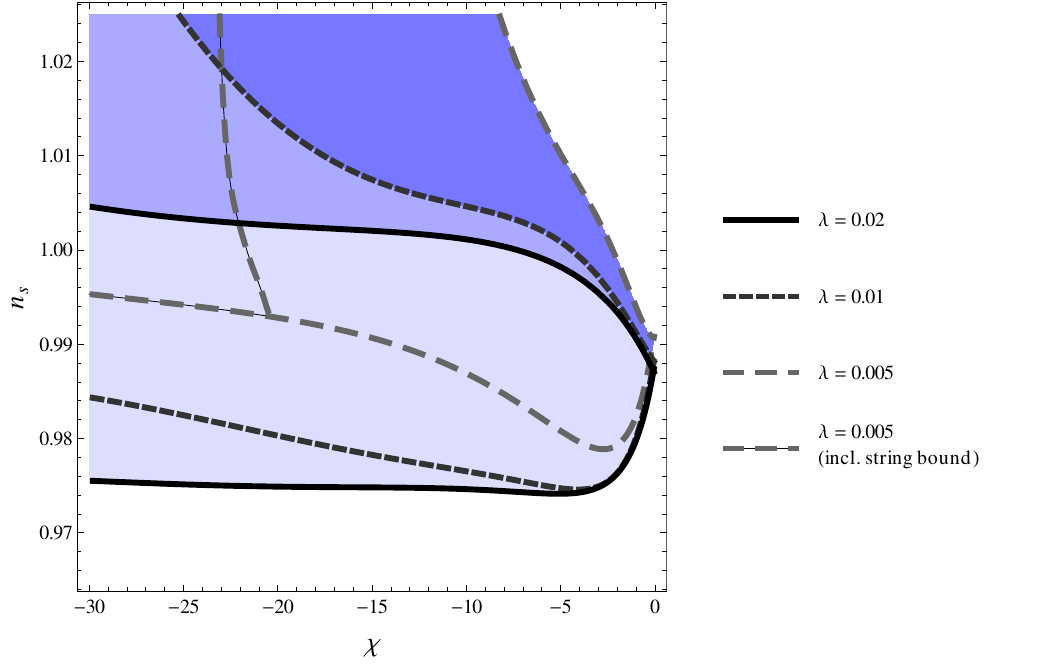}
\caption{Possible values of the spectral index $n_s$ as a function of $\chi$. The shaded region bounded by a curve with a given stroke style shows the range of possible $n_s$ values achieved by varying the inflationary trajectory for a given value of $\lambda$, while constraining the corresponding values of the amplitude to the 3-sigma range of the observed value $A_s^0$. For $\lambda = 0.005$, the region to the top left, bounded by the gray solid-dashed curve, is in accordance with the cosmic string bound.
}
\label{fig_ns_normalized}
\end{figure}


For fixed values of the parameters $\xi$ and $\lambda$, the normalization
condition, cf.\ Eq.~\eqref{eq_As0}, can be used to eliminate the parameter
$\sigma_f$, which we introduced in Section \ref{subsec_twovssingle}
to distinguish between the different inflationary trajectories.
According to Fig.~\ref{fig_ns_P_2field}, it is for instance possible to
find for $\sqrt{\xi} = 4.3 \times 10^{15}\,\textrm{GeV}$ and $\lambda = 4 \times 10^{-3}$
and for each $\chi$ value below $\chi \simeq -14.5$ one particular $\sigma_f$,
i.e.\ one inflationary trajectory such that $A_s = A_s^0$.
It is important to note that it is only these sets of parameter values,
which are compatible with the normalization condition, that we are
allowed to consider when asking for the range of viable $n_s$ values predicted by our model.


In order to determine this range of admissible $n_s$ values,
we perform a numerical scan of the parameter space and record $n_s$
for all values of the parameters $\xi$, $\lambda$, $\chi$ and $\sigma_f$ that yield
an amplitude $A_s$ within the 3-sigma range of the best-fit value $A_s^0$.
Fig.~\ref{fig_ns_normalized} presents the results of this analysis for three representative
values of the coupling constant, $\lambda = (5, 10, 20) \times 10^{-3}$, while keeping $g^2 = 1/2$ and $q = 2$.
For each $\lambda$ value, we vary $\chi$ between $-30$ and $0$
and $\sigma_f$ between $0$ and $\sigma_f^0$, where $\sigma_f^0$
is a function of $\chi$, cf.\ Eq.~\eqref{eq_sigma_f}.
Furthermore, for each $\lambda$ value, we vary $\xi$ within a small interval,
so that we cover the entire region in parameter space where the amplitude
 comes out close to the best-fit value $A_s^0$.
The lower boundaries of these intervals roughly coincide with the
respective $\xi$ values one would need in the case of single-field inflation to obtain
the correct amplitude, i.e.\ they lie on the solid blue curve in the equivalent of Fig.~\ref{fig_cobe_strings} for $\chi = -30$.
This is due to the decrease in the amplitude with decreasing $|\chi|$ as well as with decreasing $\sigma_f/\sigma_f^0$, c.f.\ Fig.~\ref{fig_P2}.
In order to compensate for this decrease one has to employ $\xi$
values in the two-field case that are a bit larger than in the single-field case.
The resulting range of $n_s$ values obtained for a given value of $\lambda$ is marked by the shaded regions bounded by curves with a given stroke style in Fig.~\ref{fig_ns_normalized}. Additionally, the solid-dashed curve marks the cosmic string bound for $\lambda = 5 \times 10^{-3}$, c.f.\ Eq.~\eqref{eq:Gmubound}, with the region to the upper left of this curve in agreement with the bound. For the two larger values of $\lambda$, the cosmic string bound is violated in the entire $\chi$-range shown.


The general trend in Fig.~\ref{fig_ns_normalized} is the same as in the
case of single-field inflation, cf. Fig.~\ref{fig_cobe_strings}:
small $\lambda$ values yield a large spectral index, while larger $\lambda$ values
give smaller $n_s$ values.
For instance, for $\lambda = 2 \times 10^{-2}$, we are able to
reach $n_s$ values below $0.98$ for nearly the entire range of $\chi$ values.
This illustrates that our model is in principle capable of generating a spectral
index of the right magnitude, while simultaneously providing the correct amplitude of
the scalar power spectrum.
An obvious problem, however, is that in order to reproduce the observed amplitude
$A_s^0$, we require quite large $\xi$ values, such that the cosmic string tension
becomes unpleasantly large.
Considering trajectories different to the $\sigma$-axis, i.e., different to the trajectory studied in Section~\ref{sec_single_field}, sharpens the tension imposed by the cosmic string bound, since the decrease in the amplitude due to the motion in $\tau$-direction forces us to go to even larger values of $\xi$ and hence larger values of $G \mu$.
%
%
%
%
Moreover, we note that among the viable values for $n_s$ for a given value of $\lambda$ and $\chi$, the
spectral index comes out smaller the closer to the $\sigma$-axis the corresponding inflationary trajectory is.
In a universe undergoing a sufficiently long period of inflation, it may however not require much fine-tuning to end up with an
inflationary trajectory running close to the $\sigma$-axis during the last $N_*$ e-folds
of inflation, cf.\ the comment below Eq.~\eqref{eq:dVeffdphi} in section~\ref{subsec_slowrolleom}.


\section{Conclusion and outlook \label{sec_conclusion}}

Superconformal symmetry is an underlying symmetry of supergravity, broken
only by fixing the value of the conformal compensator field, which
generates the kinetic term of the gravitational field. It can also serve as a guideline 
for coupling
matter fields to supergravity. The resulting supergravity models have 
several intriguing features. There is a Jordan frame where the Lagrangian
takes a particularly simple form, closely resembling global supersymmetry.
Furthermore, contrary to canonical supergravity, the scalar potential does
not contain factors which grow exponentially at large field values, which
keeps supergravity corrections to scalar masses under control. As we have seen,
a Fayet-Iliopoulos term can be introduced analogously to the kinetic
term of the graviton by making use of the conformal compensator field.

In this paper, we study hybrid models of inflation with superconformal
symmetry. As we show, the inflaton acquires a large tachyonic mass in
F-term hybrid inflation, which therefore is not viable.
On the contrary, D-term hybrid inflation is consistent with superconformal
symmetry. Allowing for an explicit symmetry breaking by a holomorphic
contribution to the K\"ahler potential involving only dimensionless parameters
\cite{Ferrara:2010yw}, one obtains a two-field inflation model.
If inflation lasted sufficiently long before the onset of the last 50 e-folds, the inflationary trajectory along the real part of the complex inflaton field becomes an attractor. For this limiting case we obtain analytic formulas 
for the amplitude of scalar fluctuations and the spectral index,
which describe the full numerical results very well. It turns out that the
spectral index can become as small as $n_s \simeq 0.96$.
For generic two-field trajectories, we calculate the resulting amplitude of the primordial power spectrum and the spectral index 
numerically.

Comparing the obtained results with current CMB data, we find that for values of the gauge coupling compatible with GUTs and after fixing the overall normalization of the primordial power spectrum to the observed value, we can identify three different regions of the parameter space. For large values of the superpotential coupling $\lambda$, we obtain a spectral index close to the current best-fit value, $n_s \simeq 0.97$. However, in this regime the model is at variance with current bounds on the cosmic string tension. 
On the other hand, for small values of $\lambda$, the cosmic string
bound can easily be fulfilled, at the price of a 2.4$\sigma$ deviation from the best-fit value for the spectral index. In the intermediate regime, the correct spectral index within 2$\sigma$ experimental uncertainty can be achieved while simultaneously fulfilling the cosmic string bound. 

Summarizing, superconformal D-term inflation can successfully account for the
primordial power spectrum, with values of the spectral index down to 
$n_s \simeq 0.96$, depending on the inflationary trajectory.
Generically, however, there is a tension with the cosmic string bound.
This might be improved by considering a more strongly coupled theory or by considering an embedding of the simple model of D-term inflation described here into a more complete setup containing additional fields, which could yield further contributions to the primordial power spectrum. It should also be noted that the bound on the cosmic string tension contains considerable theoretical uncertainties; a better understanding of the related phenomena is necessary before parameter regions in conflict with this bound can be ruled out with certainty.

\vspace{15mm}
\noindent
{\bf\large Acknowledgments}\\
The authors thank M.~Hindmarsh, T.~Jones, R.~Kallosh, J.~Louis, A.~Westphal and T.~Yanagida 
for helpful discussions.
This work has been supported by the German Science Foundation (DFG) within 
the Collaborative Research Center 676 ``Particles, Strings and the Early
Universe''.


\end{document}